# Multivariate forecast of winter monsoon rainfall in India using SST anomaly as a predictor: Neurocomputing and statistical approaches

Goutami Chattopadhyay[1], Surajit Chattopadhyay[2*], Rajni Jain[3]

[1]Formerly, Department of Atmospheric Sciences, University of Calcutta, Kolkata 700 019, India, E-mail: goutami15@yahoo.co.in

[2]Department of Computer Application, Pailan College of Management and Technology, Bengal Pailan Park, Kolkata 700 104, India Email: surajit_2008@yahoo.co.in

[3]Centre for Agricultural Economics and Policy Research, Pusa, DPS Marg, New Delhi 110012, India
Email: jainrajni@hotmail.com
* Author for correspondence

**Abstract**

In this paper, the complexities in the relationship between rainfall and sea surface temperature (SST) anomalies during the winter monsoon (November-January) over India were evaluated statistically using scatter plot matrices and autocorrelation functions. Linear as well as polynomial trend equations were obtained and it was observed that the coefficient of determination for the linear trend was very low and it remained low even when polynomial trend of degree six was used. An exponential regression equation and an artificial neural network with extensive variable selection were generated to forecast the average winter monsoon rainfall of a given year using the rainfall amounts and the sea surface temperature anomalies in the winter monsoon months of the previous year as predictors. The regression coefficients for the multiple exponential regression equation were generated using Levenberg-Marquardt algorithm. The artificial neural network was generated in the form of a multiplayer perceptron with sigmoid non-linearity and genetic-algorithm based variable selection. Both of the predictive models were judged statistically using the Willmott's index, percentage error of prediction, and prediction yields. The statistical assessment revealed the potential of artificial neural network over exponential regression.

## 1.    Introduction

The tropical Indian Ocean is influenced by two contrasting air masses (continental and marine) associated with the Indian monsoon system (Li and Ramanathan, 2002). The summer monsoon (southwest monsoon) usually starts by the end of May and continues until November. The winter monsoon (northeast monsoon) commences toward the end of November and continues until April. During summer monsoon, winds are mainly southwesterly or westerly and during winter monsoon, the prevailing winds are mostly northeasterly. The southwest (summer) and the northeast (winter) monsoons influence weather and climate between $30^0$N and $30^0$S over the African, Indian, and Asian land-masses (Reddy and Salvekar, 2003). Accurate long lead prediction of monsoon rainfall can improve planning to mitigate the adverse impacts of monsoon variability and to take advantage of the beneficial conditions (Reddy and Salvekar, 2003). The variability in the monsoon rainfall depends heavily upon the sea surface temperature (SST) anomaly over the Indian Ocean (Clark *et al*, 2000). Several authors, for example Nicholls (1983), Branstator (1985), Clark *et al* (2000), Lau and Weng (2001), Guo *et al* (2002), Barsugli and Sardeshmukh (2002), Hartmann *et al* (2007), have discussed the association between SST anomaly and rainfall. The SST anomalies cause local changes in the low-level temperatures, winds, and humidity, usually leading to precipitation anomalies in the vicinity of the SST anomaly (Barsugli and Sardeshmukh, 2002). Gradients of SST within the oceans are important in determining the location of precipitation over the Tropics, including the monsoon regions (Clark *et al*, 2000).

As the extra tropical circulation anomalies display energy dispersion away from the region of anomalous tropical convection, they are interpreted as a Rossby wave response to the latent heat release associated with the tropical convection (Ferranti *et al*., 1990). In regions of anomalous tropical heating, there is a dynamical response with anomalous large-scale ascent and upper tropospheric divergence, which acts as a Rossby wave source (Sardeshmukh and Hoskins, 1988) for extra tropical waves. Conversely, in regions of reduced convection and anomalous cooling, the tropical responses are the anomalous descent and upper-tropospheric convergence (Matthews *et al*, 2004). Positive (negative) sea surface temperature (SST) anomalies lead the enhanced (suppressed) MJO convection by approximately 10–12 days (a quarter cycle), consistent with the atmosphere responding to the ocean forcing (Matthews, 2004). The SST anomalies themselves have been simulated in thermo dynamical ocean models as the response to the observed anomalous surface fluxes of latent heat and short-wave radiation (Matthews, 2004; Shinoda and Hendon, 1998) without the need to a mass flux approach and closed on buoyancy. SST anomalies has been addressed by Woolnough *et al*. (2001), who examined the equilibrium response to an idealized, eastward-propagating equatorial SST dipole anomaly in an "aquaplanet" atmospheric general circulation model (AGCM).

The southwest (summer) and the northeast (winter) monsoons influence weather and climate between $30^0$N and $30^0$S over the African, Indian and Asian landmasses. Sea Surface Temperature (SST) anomalies influence the atmosphere by altering the flux of latent heat and sensible heat from the ocean (Holton, 1972). The efficiency of such anomalies in exciting global scale responses depends on their ability to generate Rossby waves. In the extra tropics SST anomalies primarily generate low-level heating and this is balanced by horizontal temperature advection. In the tropics, positive SST anomalies are associated with enhanced convection and the resulting heating is balanced by adiabatic cooling. SST anomalies also play an important role in producing rainfall (Reddy and Salvekar, 2003).

El-Niño-Southern Oscillation (ENSO) is a coupled Ocean-atmosphere phenomenon that has worldwide impact on climate in general and Indian monsoons in particular. The oscillations in wind stress owing to the Southern Oscillation are associated with changes in the circulation of the ocean and the SST anomaly that are referred to as El-Nino. This term was originally applied to a warming of the coastal waters of Peru and Ecuador, which occurs annually near Christmastime. ENSO's maximum SST anomalies occur in the equatorial eastern and central Pacific, ENSO affects the global ocean (Wang *et al*, 2006). Outside the tropical Pacific significant ENSO related SST anomalies are found over many places, such as in the tropical North Atlantic, the tropical Indian Ocean, the extratropical North and South Pacific, and the South China Sea. The strength of ENSO is measured by two principle indices namely Southern Oscillation Index (SOI) and SST index over Nino-3.4 region, which refers to, the anomalous SSTs within the region bounded by $50^oN$-$50^oS$ and $170^oE$-$120^oW$. Warm ENSO episodes are characsied by increased number and intensity of tropical storms over the Bay of Bengal and hence enhanced winter monsoon rainfall.

The long-recognized negative correlation between Indian monsoon rainfall and ENSO (Webster and Palmer, 1977) in which a weak (strong) monsoon is related to a warm (cold) event through an anomalous Walker cell driven by tropical east Pacific sea surface temperature (SST) anomalies, has weakened rapidly since the late 1970s (Chang *et al*, 2001). This weakened relationship is defined by the correlation between June and September all-India rainfall and Niño-3 (5°S–5°N, 150°–90°W) SST (Chang *et al*, 2001). A handful of significant studies are available where the relationship between SST and rainfall has been studied (e.g. Uvo *et al*, 1989; Nobre and Srukla, 1996). Chongyin (1990) investigated the relationship between El-Nino event and winter monsoon over south east Asia.

Present paper attempts to develop a predictive model for winter monsoon over India using SST anomaly as one of the predictors. Some significant studies in the field of winter-monsoon over East Asia and its association have been reviewed. The literatures include Zhang *et al* (1997), Wen *et al* (2000), and Quiaoping *et al* (2007). However, limited studies could be identified where prediction of winter-monsoon over India has been discussed. Finally, we decided to generate a predictive model for winter-monsoon over India. From the above discussion and the literatures mentioned above we felt the necessity of incorporating SST anomaly as a predictor. Since artificial neural network (ANN) is being attempted in several papers dealing with complex non-linear meteorological events (e.g. Elsner and Tsonis, 1992), we decided to attempt an ANN in this problem. The subsequent sections are organized as follows: Section 1.1 discusses some literatures in the application of ANN in rainfall prediction. Section 2 describes the inputs to the model and examines the structure of the time series under consideration. Section 3 generated a non-linear multiple regression model in the form of exponential regression. Section 4 develops and describes the ANN model and Section 5 makes a statistical assessment of the predictions from regression and ANN. The conclusion is presented in Section 6.

## 1.1 Artificial Neural Network in rainfall prediction – a brief overview

Accurate rainfall predictions are essential for planning day-to-day activities. Several papers are available where the rainfall time series have been dealt with statistically (Wilks, 1991; Ward and Folland, 1991; Bell and Suhasini, 1994). Present paper endeavors to develop an Artificial Neural Network (ANN) model to forecast average winter shower in India. The ANN technique is widely accepted as a potentially useful way of modeling complex non-linear and dynamic systems for which a large amount of data is used (Hsieh and Tang, 1998; Hornik, 1991). It is particularly useful, where the underlying physical processes or relationships are not fully understood or may display chaotic properties (Sivakumar, 2000). Conventional weather forecasting models are highly data specific and based on complex and expensive to maintain mathematical models that are built many months in advance of the events. ANN offer real prospects for an effective, more flexible, less assumption dependent adaptive methodology well suited for modeling weather forecasting which by its nature are inherently complex because of non-linearity and chaotic effects (Maqsood *et al*, 2002). The ANN is based on biological neural systems (Kartalopoulos, 2000). It is highly vital with respect to underlying data distribution and no assumptions are made about relationships between variables. The basic unit of ANN is the artificial neuron, which stimulates the four basic functions of natural neurons. The artificial neurons are much similar than biological neuron. The input layers consist of neuron that received input from the external environment. The output of the system layer consists that communicate the output of the system to the user of the external environment. There are usually numbers of hidden layers between these two layers. To determine the number of hidden neurons the network should have perform its best, are often left out to the method trial and error. Usually the available data are separated into training set and test set. The optimal weights are obtained by using 'Back Propagation'. The quality of the prediction is obtained from the performance with the test set of data. The weights are determined by iteration to produce the lowest error in the output. Initial weights are randomly assigned in subsequent iterations, individual weights are incrementally adjusted to reduce error (Kamarthi and Pittner, 1999).

Hu (1964) initiated the execution of ANN in weather forecasting. Since the last few decades, ANN has opened up new avenues to the forecasting task involving atmosphere related phenomena (Gardner and Dorling, 1998; Hsieh and Tang, 1998). Michaelides *et al* (1995) compared the performance of ANN with multiple linear regressions in estimating missing rainfall data over Cyprus. Kalogirou *et al* (1997) implemented ANN to reconstruct the rainfall time series over Cyprus. Lee *et al* (1998) applied Artificial Neural Network in rainfall prediction by splitting the available data into homogeneous subpopulations. Wong *et al* (1999) constructed fuzzy rule bases with the aid of SOM and Back propagation neural networks and then with the help of the rule base developed predictive model for rainfall over Switzerland using spatial interpolation. Application of ANN in forecasting monsoon rainfall over India is not very new. We have surveyed the literature and could get some significant papers in this direction. Venkatesan *et al* (1997), Sahai (2000), Chattopadhyay (2007), Guhathakurata (2008) implemented ANN based methodologies in prediction of summer-monsoon rainfall over India.

## 2. Autocorrelation structure of the time series

The present paper deals with the monthly average winter shower data and the SST anomaly data from 1950 to 1998. The rainfall and SST anomaly data have been collected from www.tropmet.res.in and http://jisao.washington.edu/. The winter shower (mm) and tropical SST anomaly ($^0C$) has been examined for their autocorrelation pattern. We have calculated autocorrelation function (ACF) (Wilks, 1995) for the average (over November-January) rainfall time series up to 25 lags and displayed in Figure 1a. This Figure shows that the ACF does not follow any specific sinusoidal pattern and the autocorrelations fall below 0.3 in numerical value. Moreover, the ACF is not decaying to 0 with increase in the lag number. These features of the ACF indicate that the time series does not have any persistence and it is not stationary in nature. The ACF of the average winter monsoon rainfall has been examined because this is the predictand of the modeling problem. The monthly rainfall amounts in November, December, and January are also examined for their ACF and similar nature is discerned (Figures 1b-d).

Similarly, the ACF are calculated for tropical SST anomaly in the said months (Figures 2a-c) and it is revealed that ACF for the SST time series are remaining at positive level up to more than 10 lags. However, the ACF is lying below 0.5 in magnitude. This indicates that despite more positive association with past values than the winter monsoon rainfall no persistence exists within the time series. Like the rainfall time series, the ACF for SST anomaly is not showing any sinusoidal pattern and hence it can be said that stationarity is not there with the SST anomaly time series. Thus, it can be said that neither the time series of the predictors (i.e. rainfall amounts and SST anomaly in November, December, and January for year $n$) nor the time series of the predictand (i.e. average winter monsoon rainfall for year $(n+1)$) exhibits stationarity. The non-linear association between the predictors and the predictand is also apparent from the widely distributed data clouds in the scatter plot presented in Figure 3.

## 3. Multiple non-linear regression model

Initially, all of the six predictors are tested over time for their trend. Linear as well as non-linear (polynomial of degree 6) trend equations (Wilks, 1995) are examined and the corresponding coefficients of determination ($R^2$) are calculated. We have presented the values of $R^2$ in Table 1, which shows that in all of the cases the values are small. This indicates the absence of any significant trend within the time series of all of the predictors. In the previous section, the non-stationarity of the time series under study has been revealed. It is also understandable from the ACF patterns that there are no seasonality and trend in the time series. Non-linear regression equations are now generated to discern the relationship between the predictors and predictands. A non-linear regression equation involving exponential terms is generated as follows (Chattopadhyay and Chattopadhyay, 2008a):

$$\hat{y} = \sum_{i=1}^{6} a_i \, exp(b_i x_i) \quad \ldots \quad \ldots \quad \ldots \quad (1)$$

Where, $a_i$, $b_i$ are the regression parameters, and $x_i (i=1,2,...,6)$ denote the six predictors i.e. rainfall in November, December, January, and the tropical SST anomalies in November, December, and January respectively for the year $n$; in the left hand side $\hat{y}$ denotes the estimated average winter monsoon rainfall in the year $(n+1)$. The regression parameters are estimated using Levenberg-Marquardt algorithm and iterating up to 500 steps. Finally, the regression parameters come out to be 0.355 $(a_1)$, 0.031 $(b_1)$, 4.705 $(a_2)$, 0.018 $(b_2)$, 2.505 $(a_3)$, 0.056 $(b_3)$, 4.730 $(a_4)$, 0.001 $(b_4)$, 3.343 $(a_5)$, 0.001 $(b_5)$, 3.480 $(a_6)$, and 0.002 $(b_6)$. The results would be analyzed statistically in the subsequent sections.

## 4. Artificial neural network model

In the present paper, the data have been transformed to $[-1,+1]$ before applying them to ANN formation. An exhaustive variable selection procedure is adapted to find a good subset of the full set of input variables created as a result of data analysis and transformation. This selection has been done by means of genetic algorithm, where the cross over probability and mutation factor has been selected as 0.7 and 1 respectively. While generating the ANN model in the form of multiplayer perceptron (MLP), using a modified cascade method together with an adaptive gradient learning rule (Lundin *et al*., 1999). The cascade mode of construction involves adding hidden nodes, one or more than one at a time, and always connecting all the previous nodes to the current node. The adaptive gradient learning rule uses back-propagated gradient information to guide an iterative line search algorithm. Sigmoid function ($f(x) = (1 + e^{-x})^{-1}$) (Widrow and Lehr, 1990) has been used as activation function for both hidden and output layer. From the entire dataset under consideration, we have chosen 70 per cent of the original data as training set. The training set formation has been done using the method of Round Robin (O'Neill and Song, 2003), which is an arrangement of choosing all elements in a group equally in some rational order, usually from the top to the bottom of a list and then starting again at the top of the list and so on. The root mean squared error (RMSE) has been used to evaluate the model. After training, the network has been validated over the entire dataset. It should be mentioned that there is no strict rule to decide the ratio of training and test cases. A survey on the ANN literatures it is found that the ratios 1:1 (Chattopadhyay and Chattopadhyay, 2008b), 7:3 (Lundin *et al*., 1999), and 3:1 (Perez *et al*., 2000, Perez and Reyes, 2001) are frequently used in ANN applications. In the present paper the approach similar to that of Lundin *et al*. (1999) has been adopted after examining the other approaches. The ANN has been validated over the entire dataset under consideration. We have started with 30 nodes in the hidden layer and after training through adaptive gradient learning, we have obtained the final structure of the MLP as 5-12-1. This implies that there are five units (it is reduced from six due to exhaustive variable selection) in the final input set, twelve nodes in the final hidden layer, and there is only one output node. Results obtained from this network architecture are explained in the subsequent section.

## 5. Statistical assessment of the predictions

In the last two sections, we have generated non-linear regression and ANN models. In the present section, the results would be analyzed. A confusion matrix has been generated to view the prediction capacity of the ANN model. A confusion matrix (Fielding and Bell, 1997) is a square matrix whose rows and columns represent the sub-ranges for the real world target and model outputs, respectively. The value in the (i,j) position of the matrix is the number of records for which the real world target output is in the ith sub-range and whose real world model output is within the jth sub-range. The confusion matrix for the ANN prediction comes out to be

| | | Actual sub ranges | | | | | |
|---|---|---|---|---|---|---|---|
| | Confusion matrix | 4.23 | 13.13 | 22.02 | 30.91 | 39.81 | Totals |
| Predicted sub ranges | 4.23 | 2 | 8 | 0 | 0 | 0 | 10 |
| | 13.13 | 2 | 10 | 3 | 1 | 0 | 16 |
| | 22.02 | 0 | 4 | 8 | 0 | 0 | 12 |
| | 30.91 | 0 | 1 | 2 | 2 | 1 | 6 |
| | 39.81 | 0 | 0 | 2 | 0 | 1 | 3 |
| | Total | 4 | 23 | 15 | 3 | 2 | 47 |

From the above matrix, it is found that for the sub range (4.23mm to 13.13mm) there are actually 23 observations and 16 predictions fall in this sub range. Thus, 69.57% accuracy is there for this sub range. For the sub range 13.13mm to 22.02mm, there are actually 15 observations and 12 predictions fall in this sub range. This means that there is 80% accuracy for this sub range. Thus, it can be interpreted that for higher rainfall accuracy there is more possibility of accurate fore cast by ANN than in the case of lower rainfall amounts. In the next step, the predictions are judged using scatter plot matrix presented as Figure 4. In this Figure it is apparent from the second panel of the first column that there is a good linear association between the actual and the predicted values of the monsoon rainfall amounts based on ANN model described above. The dense data cloud around the linear trend line indicates a good linear association between the actual and the predicted values. However, there are some values deviated significantly from the trend line. Thus, there are some cases where the prediction performance of the ANN is not satisfactory. Now, the predictions from ANN are judged case wise over the entire validation set. Line diagram presented in Figure 5 shows the comparison between actual and predicted rainfall amounts for the validation cases. A visual inspection over Figure 5 reveals that for the validation cases 1, 2, 3, 4, 6, 7, 9, 10, 12, 14, 19, 22, 23, 25, 31, 32, 33, 38, 41, 42, 44, and 45 there is significant closeness between actual and predicted rainfall amounts. Computing the per centage of errors in the validation cases we find that if maximum 15% error of prediction is allowed, then prediction yield is 0.30. If 20%, 25%, and 30% errors of prediction are allowed, then the prediction yields are 0.45, 0.55, and 0.62 respectively. Now we consider the predictions from the exponential regression presented in equation (1). Observing the scatter plot presented in Figure 4, we find that there in a near-linear association between actual and regression-predicted values of the rainfall. However, a visual inspection reveals from the third row and first column of the matrix that the linearity is less in this case than that in the case of ANN. To test the results statistically we calculate the Willmott's index (WI) for the regression as well as ANN. The Willmott's index is given by (Willmott, 1982)

$$\text{WI} = 1 - \left[ \sum_{i=1}^{N}(P_i - O_i)^2 / \sum_{i=1}^{N}\left(|P_i - \overline{O}| + |O_i - \overline{O}|\right)^2 \right] \qquad \ldots \qquad (2)$$

Where, $O_i$, $P_i$, and $\overline{O}$ denote the observed value in the ith case, predicted value in the ith case, and mean observed value respectively. The utility of WI in judging the predictions has been discussed thoroughly in Chattopadhyay and Chattopadhyay (2008). In the present case, the WI for non-linear regression and ANN come out to be 0.672 and 0.724 respectively. As the higher value of WI indicates better predictive model, it can be said that the ANN based predictive model performs better than non-linear regression. Further, we calculate percent error of the prediction (PE) given by (Chattopadhyay, 2007)

$$\text{PE} = \frac{\langle |P_i - O_i| \rangle}{\langle O_i \rangle} \times 100 \qquad \ldots \qquad (3)$$

Where, $\langle \ \rangle$ implies the average over the whole test set. The predictive model is identified as a good one if the PE is sufficiently small. In the present case, PE for non-linear regression and ANN come out to be 30.96 and 27.16 respectively. This further proves that ANN is performing better than regression in predicting winter monsoon rainfall. Like ANN, the prediction yields are also calculated for the non-linear regression based prediction. It is found that if 15%, 20%, 25%, and 30% errors of prediction are allowed, then the prediction yields are 0.23, 0.30, 0.45, and 0.51 respectively, and each of the prediction yields is less than the corresponding values in the case of ANN. This further proves the better prediction potential of ANN than regression. For better viewing, the values of the statistical parameters are presented in Table 2.

## 5. Conclusion

The present paper dealt with the prediction of winter-monsoon rainfall time series over India. The purpose was to predict the average winter-monsoon rainfall using six predictors. To do the same, we considered the rainfall amounts and the sea surface temperature anomalies for the winter monsoon months of a given year to predict the average winter monsoon rainfall of the next year. Consequently, we had six predictors and one predictand. Before generation of the model, we discussed several issues associated with rainfall and its relationship with sea surface temperature anomalies. Subsequently we surveyed the literatures dealing with the applicability of artificial neural network to the forecasting of rainfall time series. Prior to develop a neural network model we first tried a predictive non-linear multiple regression equation after thoroughly investigating the autocorrelation structures and scatter plot matrices for the predictors and the predictand. The highly non-linear trends and correlations as well as absence of any serial correlation in the time series under study prompted us to go for a non-linear regression equation in the form of exponential regression. After fixing the regression coefficients and the regression constant by means of Levenberg-Marquardt algorithm we calculated some statistical parameters to assess the goodness of the prediction made by the exponential regression. Afterwards, we developed the artificial neural network model multiplayer perceptron using a modified cascade method together with an adaptive gradient learning rule. After training and validating through extensive variable selection procedure final structure of the multiplayer perceptron was obtained as 5-12-1. The predictions from the neural net were also judged using the same statistical parameters, i.e., Willmott's index, percentage error of prediction and prediction yields. All the statistical parameters revealed the supremacy of artificial neural network over exponential regression in predicting the winter-monsoon rainfall over India. Simultaneously, scatter plots were also made between the observed and predicted values for both neural network and multiple regression. In the scatter plots, more linearity was observed in the case of neural network that in the case of regression.

**References**

Barsugli, J.J. and Sardeshmukh, P.D., 2002, Global atmospheric sensitivity to tropical SST anomalies throughout the Indo-Pacific basin, *Journal of Climate*, **15**, 34427-3442.

Branstor, G., 1985, Analysis of general circulation model sea-surface temperature anomaly simulations using a linear model. Part I: Forced solutions, *Journal of Atmospheric Sciences*, **42**, 2225-2241.

Bell, T. L. and Suhasini, R., 1994, Principal modes of variation of rain-rate probability distribution. *Journal of Applied Meteorology*, **33**, 1067-1078.

Clark, O.C., Cole, J.E. and Webster, P.J., 2000, Indian Ocean SST and Indian summer monsoon rainfall: predictive relationships and their decadal variability. *Journal of Climate* **14**: 2503-2519

Chattopadhyay, S., 2007, Feed forward Artificial Neural Network model to predict the average summer-monsoon rainfall in India, *Acta Geophysica*, **55**, 369-382

Chattopadhyay, S. and Chattopadhyay, G., 2008a, Comparative study among different neural net learning algorithms applied to rainfall time series, *Meteorological Applications*, **15,** 273-280.

Chattopadhyay, S. and Chattopadhyay, G., 2008b, Identification of the best hidden layer size for three layered neural net in predicting monsoon rainfall in India, *Journal of Hydroinformatics*, **10**, 181-188.

Chang, C.P., P. Harr, and J. Ju, 2001: Possible Roles of Atlantic Circulations on the Weakening Indian Monsoon Rainfall–ENSO Relationship. *J. Climate*, **14**, 2376–2380.

Chongyin, L., 1990, Interaction between anomalous winter monsoon in East Asia and El Nino events, Advances in Atmospheric Sciences, 7, 36-46.

Ferranti, L., Palmer, T. N., Molteni, F. and Klinker, E. (1990) Tropical-extratropical interaction associated with the 30- 60 day oscillation and its impact on medium and extended range prediction. *J. Atmos. Sci.,* **47** : 2177-2199.

Fielding, A.H. and Bell, J.F., 1997, A review of methods for the assessment of rediction errors in conservation presence/absence models, *Environmental Conservation,* **24**, 38-49.

Gardner MW, Dorling SR (1998) Artificial Neural Network (Multilayer Perceptron)- a review of applications in atmospheric sciences. *Atmospheric Environment* **32**: 2627-2636

Guhathakurta, P., 2008, Long lead monsoon rainfall prediction for meteorological sub-divisions of India using deterministic artificial neural network model, *Meteorology and Atmospheric Physics*, 101, 93-108.

Guo Y, Zhao Y, Wang J. 2002. Numerical simulation of the Relationships between the 1998 Yangtze river valley flood and SST anomalies. *Advances in Atmospheric Sciences* **19**: 391–404.

Hartmann, H., Becker, S. and King, L., 2007, Predicting summer rainfall in the Yangtze River basin with neural networks, *International Joutrnal of Climatology* (on-line first), DOI: 10.1002/joc.1588

Hsieh, W. W. and Tang, T. (1998) Applying Neural Network Models to Prediction and Data Analysis in Meteorology and Oceanography. *Bulletin of the American Meteorological Society* **79**: 1855-1869

Holton, JR, (1972) An Introduction to Dynamic Meteorology. Academic Press, San Diego.

Hornik, K., (1991) Approximation capabilities of multilayer feedforward networks. *Neural Networks,* **4**: 251-257

Hsieh, W. W. & Tang, T. (1998) Applying Neural Network Models to Prediction and Data Analysis in Meteorology and Oceanography. *Bulletin of the American Meteorological Society* **79**: 1855-1869.

Kartalopoulos SV (2000) Understanding Neural Networks and Fuzzy Logic- Basic Concepts and Applications, Prentice Hall, New-Delhi

Kamarthi SV, Pittner S (1999) Accelerating neural network training using weight extrapolation. *Neural Networks* **12**:1285-1299

Kumar, B., (2005) Impact of ENSO on Winter Monsoon Rainfall over South India. *Geophysical Research Abstract*, **7**.

Lau, K.M. and Weng, H., 2001, Coherent modes of global SST and summer rainfall over China: an assessment of the regional impacts of the 1997–98 El Nino, *Journal of Climate*, **14**, 1294-1308.

Lee, S., Cho, S. and Wong, P.M. (1998) Rainfall prediction using Artificial Neural Network. *Journal of Geographic Information and Decision Analysis* **2**: 233-242

Mart´ýn del Br´ýo B, and Serrano Cinca, C (1993) Self-organizing neural networks for analysis and representations of data: Some financial cases. *Neural Computing and Applications* **1**: 193–206

Maqsood I, Muhammad RK, Abraham A (2002) Neurocomputing Based Canadian Weather Analysis. Computational Intelligence and Applications. Dynamic Publishers Inc., USA: 39–44.

Matthews, A. J., Hoskins B. J. and Masutani, M. (2004) The Global Response to Tropical Heating in the Madden-Julian Oscillation during Northern Winter, *Q.J. Royal Met Society,* **130**, 1-20.

Matthews, A. J. (2004) Atmospheric response to observed intraseasonal tropical SST anomalies. *Geophysical Research Letters* **31**, doi:10.1029/ 2004GL020474

Nobre, P., and Srukla, J.,1996, Variations of sea surface temperature, wind stress, and rainfall over the tropical Atlantic and South America. *Journal of Climate*, **9**, 2464–2479.

Nicholls, N., 1983: Predicting Indian monsoon rainfall from sea surface temperature in the Indonesia-north Australia area. *Nature,* **306,** 576–577.

Perez P, Trier A, Reyes J (2000) Prediction of PM2.5 concentrations several hours in advance using neural networks in Santiago, Chile. *Atmospheric Environment* **34**: 1189-1196

Perez P, Reyes J (2001) Prediction of particulate air pollution using neural techniques. *Neural Computing and Application* **10**:165-171

Qiaoping, L., Yihui, D., Wenjie , D and Guanhua, Y ., 2007, A numerical study on the winter monsoon and cold surge over East Asia, *Advances in Atmospheric Sciences*, **24**, 664-678.

Reddy, P.R.C. and Salvekar, P.S. (2003) Equatorial East Indian Ocean sea surface temperature: A new predictor for seasonal and annual rainfall. *Current Science* **85**: 1600-1604.

Sahai, A. K., Soman, M. K. and Satyan, V., 2000, All India summer monsoon rainfall prediction using an artficial neural network, Climate Dynamics, 16:291-302.

Sardeshmukh, P. D. and Hoskins, B. J (1988) The generation of global rotational flow by steady idealized tropical divergence. *J. Atmos. Sci.,* **45,** 1228-1251

Smith, T. M., R. W. Reynolds, R. E. Livezey, and D. C. Stokes, 1996: Reconstruction of historical sea surface temperatures using empirical orthogonal functions. *J. Climate,* **9:** 1403-1420.

Sivakumar, B. (2000) Chaos theory in hydrology: important issues and interpretations. *Journal of Hydrology* **227**: 1–20.

Shinoda, T., and H. H. Hendon (1998), Mixed layer modeling of intraseasonal variability in the tropical Pacific and Indian Oceans, *J. Clim.,* **11**, 2668–2685.

Uvo, C.B., C.A. Repelli, S.E. Zebiak, and Y. Kushnir, 1998: The Relationships between Tropical Pacific and Atlantic SST and Northeast Brazil Monthly Precipitation. *J. Climate*, **11**, 551–562.

Venkatesan, C. ,. Raskar, S. D., Tambe, S. S., Kulkarni, B. D. and Keshavamurty, R. N., 1997, Prediction of All India Summer Monsoon Rainfall Using Error-back-propagation Neural Networks, *Meteorology and Atmospheric Physics*, **62**, 225-240.

Webster P. J., and T. N. Palmer, 1997: The past and the future of El Niño. *Nature*, **390**, 562–564.

Ward, M. N. and Folland, C., 1991, Prediction of seasonal rainfall in the north Nordeste of Brazil using eigenvectors of sea-surface temperature. *International Journal of Climatology*, **11**, 711-745.

Wen, C., Graf, H.F. and Ronghui, H., 2000, The nnterannual variability of East Asian winter monsoon and its relation to the summer monsoon, *Advances in Atmospheric Sciences*, **17,** 48-60.

Wilks DS (1995) Statistical Methods in Atmospheric Sciences, Academic Press, USA.

Wilks DS (1991) Representing serial correlation of meteorological events and forecasts in dynamic decision – analytic models. *Monthly Weather Review* **119**: 1640-1662

Widrow B, Lehr MA (1990) 30 years of Adaptive Neural Networks: Perceptron, Madaline and Backpropagation, *Proceedings of IEEE* **78**: 1415-1442

Wang C., Wang W., Wang D., Wang Q., (2006). Interannual variability of the South China Sea associated with El Nino. *Journal of Geophysical Research* **III** : 1-19

Woolnough, S. J., *et al*. (2000), The relationship between convection and sea surface temperature on intraseasonal timescales, *J. Clim.*, **13**, 2086–2104.

Xue, Y. and Shukla, J., (1997) Model Simulation of the Influence of Global SST Anomalies on Sahel Rainfall. *Monthly Weather Review,* **126**: 2782-2792

Zhang, Y., Sperber, K. R., Boyle, J. S., Dix, M., Ferranti, L., Kitoh, A., Lau, K. M., Miyakoda, K., Randall, D., Takacs, L. and Wetherald, R., 1997, East Asian winter monsoon: results from eight AMIP models, *Climate Dynamics,* **13,** 797-820.

Table – 1. Coefficient of determination ($R^2$) of linear and polynomial trends

| Pairs of parameters | $R^2$ for linear trend | $R^2$ for polynomial trend |
|---|---|---|
| November-SST | 0.0029 | 0.1725 |
| December-SST | 0.0026 | 0.1592 |
| January-SST | 0.0052 | 0.1241 |
| November-shower | 0.7434 | 0.7484 |
| December-shower | 0.0822 | 0.1574 |
| January-shower | 0.024 | 0.0466 |

Table – 2. Values of different statistical parameters used to asses the prediction potential of regression and ANN based models

| Models | Willmott's index | Percentage error of prediction | Prediction yield (for 15% error) | Prediction yield (for 20% error) | Prediction yield (for 25% error) | Prediction yield (for 25% error) |
|---|---|---|---|---|---|---|
| Regression | 0.67 | 30.96 | 0.23 | 0.30 | 0.45 | 0.51 |
| ANN | 0.72 | 27.16 | 0.30 | 0.45 | 0.55 | 0.62 |

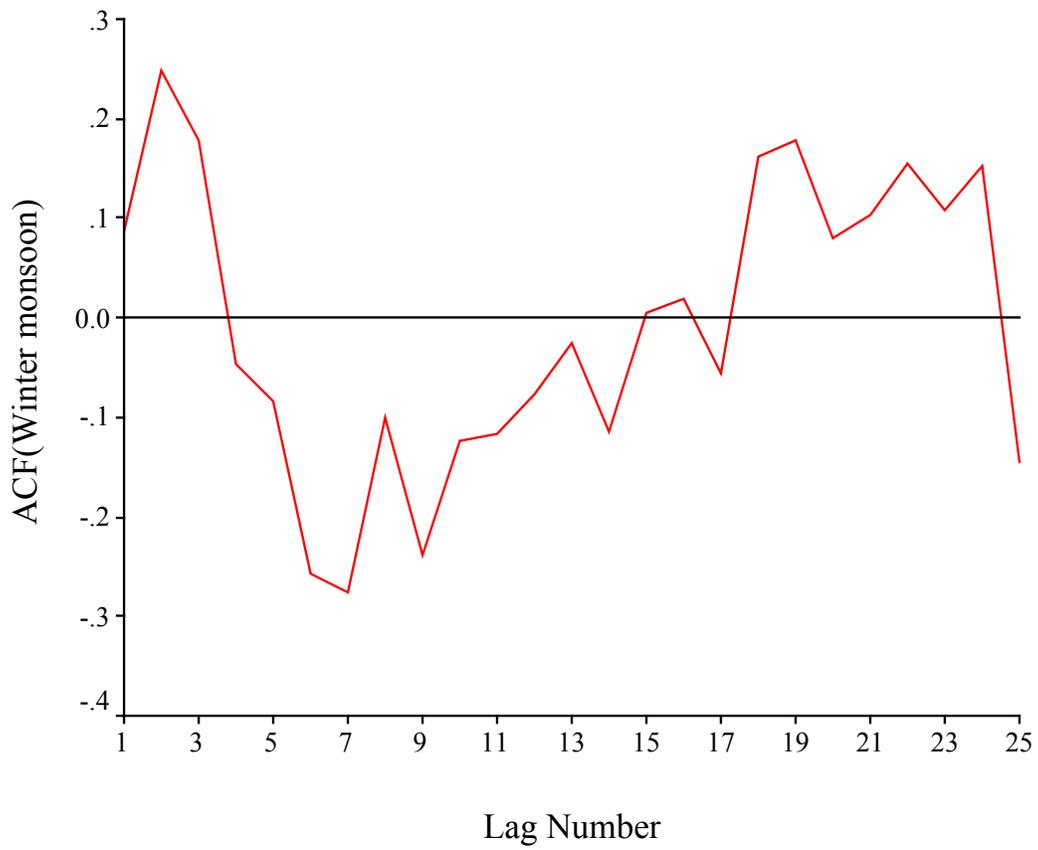

Figure 1a. Autocorrelation function (ACF) for rainfall averaged over November, December, and January in India.

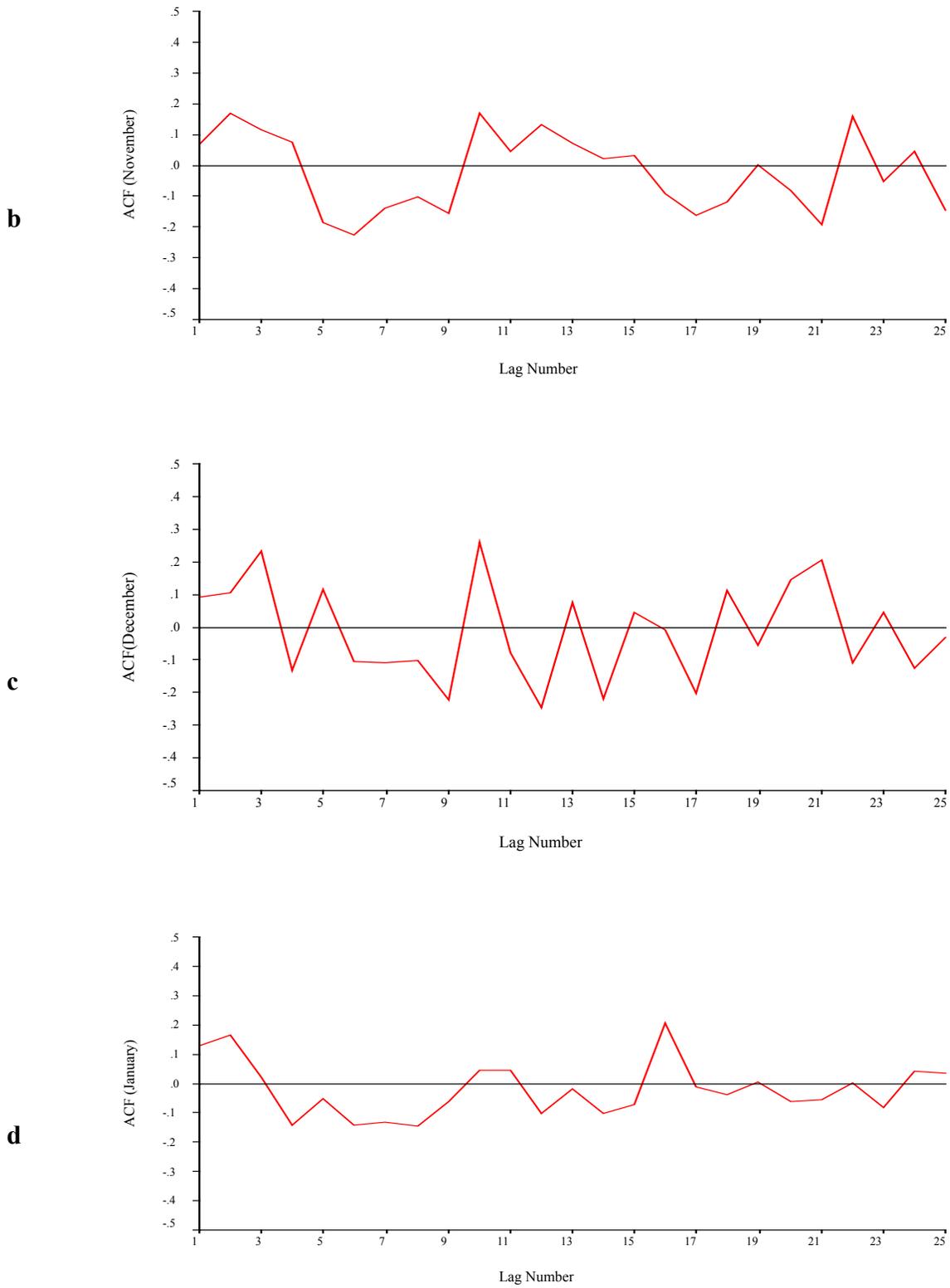

Figure 1. Autocorrelation function (ACF) for rainfall during November (b), December (c), and January (d) over India.

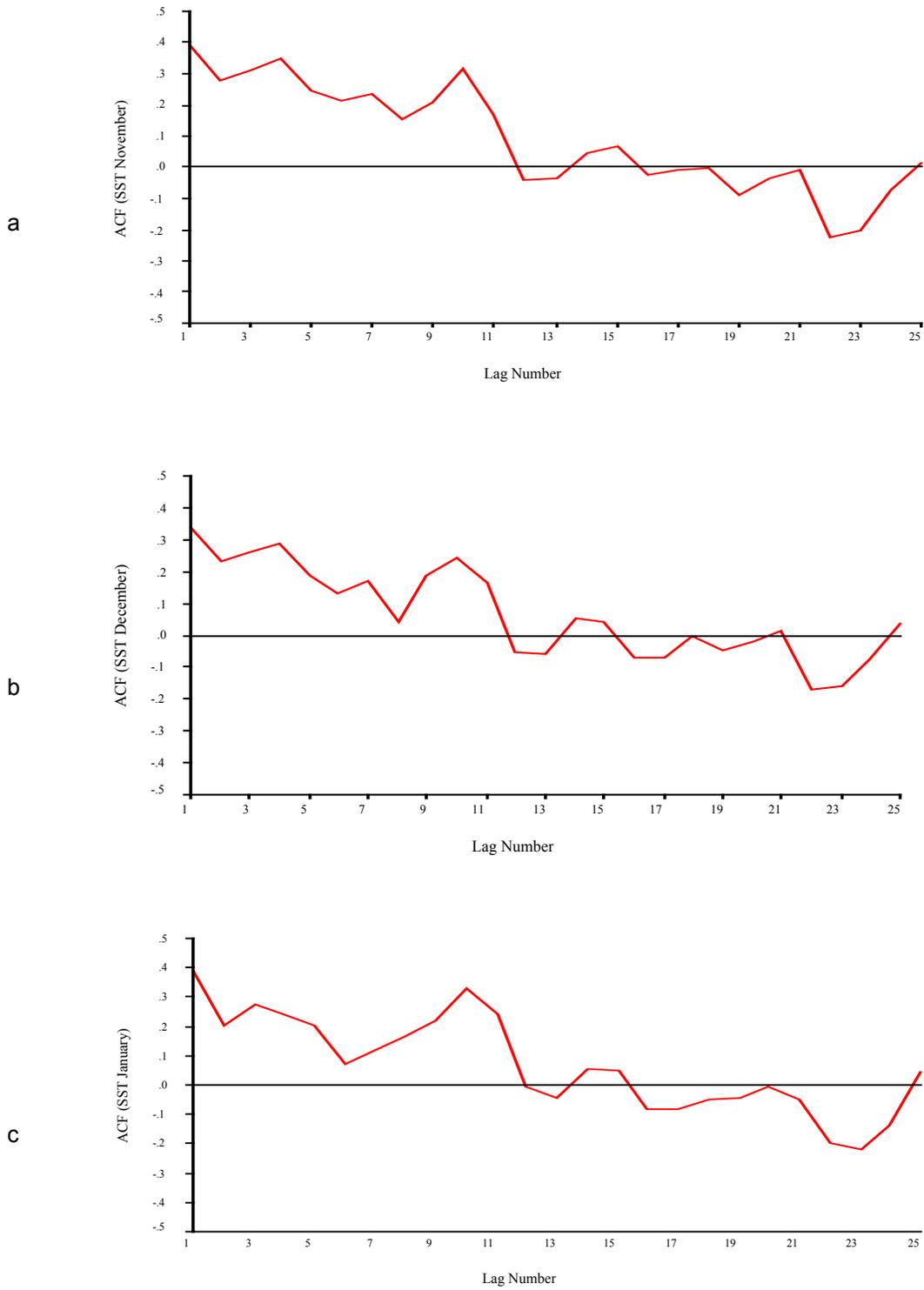

Figure 2. Autocorrelation function (ACF) for tropical sea surface temperature (SST) anomaly during November (a), December (b), and January (c).

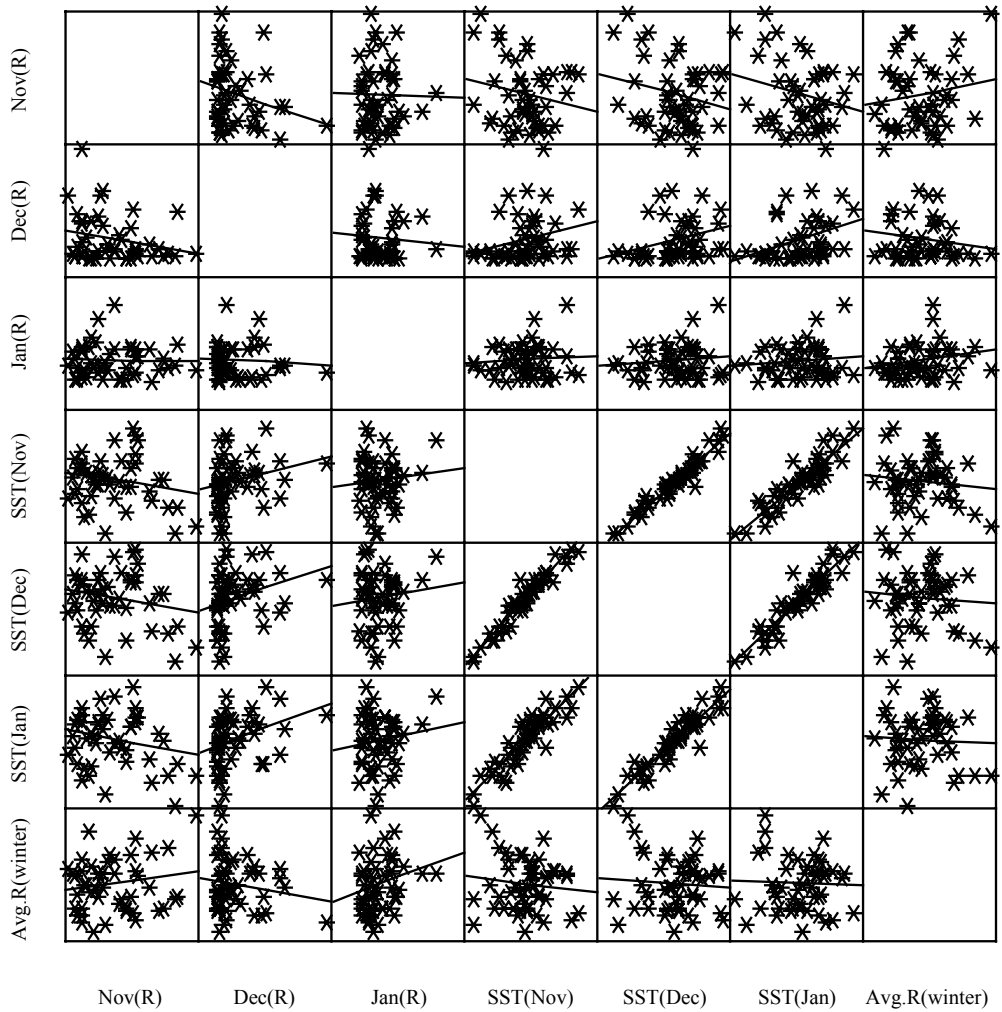

Figure 3. Scatterplot showing the association between pairs of data considered in the study.

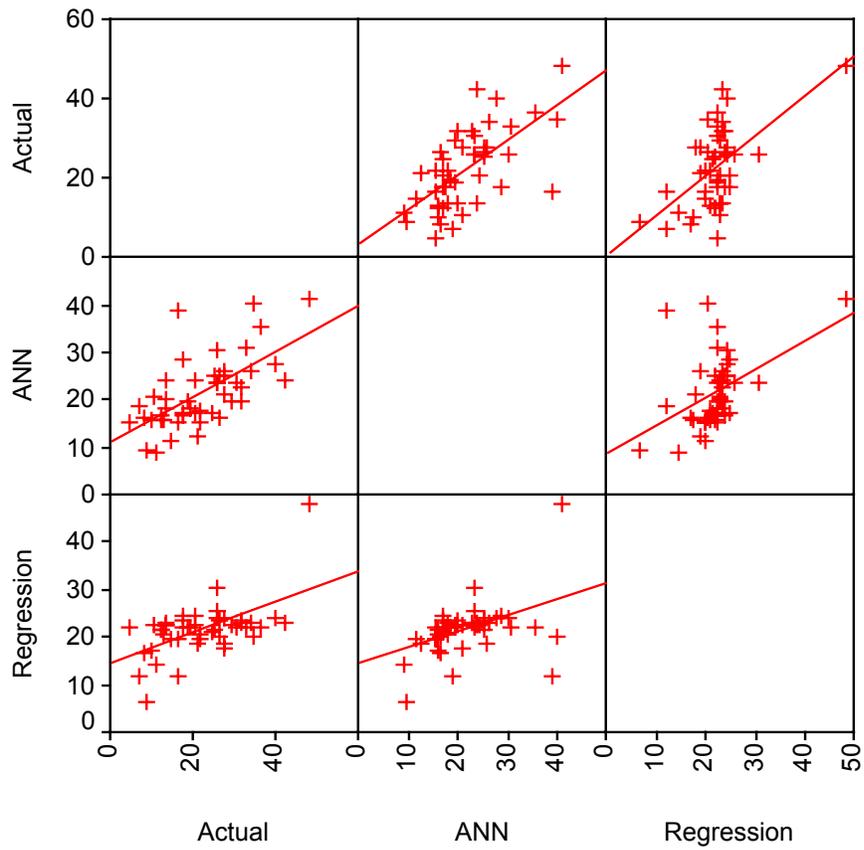

Figure 4. Scatter plot showing the associations between actual rainfall amounts and those predicted by the artificial neural network (ANN) and non-linear regression.

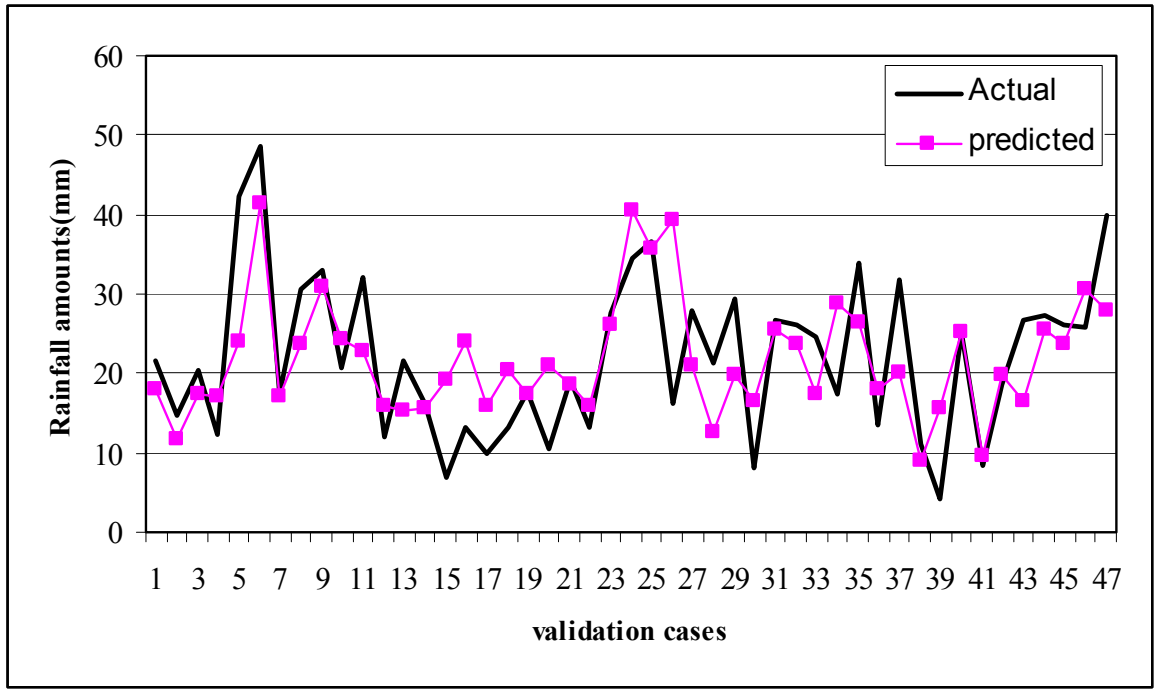

Figure 5. Line diagram of the actual and predicted rainfall amounts for the validation cases of the ANN.